\documentstyle[12pt,fleqn]{article}
\topmargin - 0.8cm

\catcode`@=11
\@addtoreset{equation}{section}
\catcode`@=12
\mathchardef\SGamma="7100

\begin{document}
\title{\vskip-1.7cm \bf  Quantum Origin of the Early
Universe and the Energy Scale of Inflation
\footnote{Talk given at the 6th Moscow Quantum Gravity Seminar,
Moscow, June 12-19, 1995. To be published in Proceedings.}}
\author{A.O.Barvinsky$^{1}$\ and A.Yu.Kamenshchik$^{2}$ }
\date{}
\maketitle
\hspace{-8mm}$^{1}${\em
Theory Department, Lebedev Physics Institute, Lebedev Research Center in
Physics, Leninsky Prospect 53,
Moscow 117924, Russia}
\\ $^{2}${\em Nuclear Safety
Institute, Russian Academy of Sciences , Bolshaya Tulskaya 52, Moscow
113191, Russia}
\begin{abstract}
Quantum origin of the early inflationary Universe from the
no-boundary and tunnelling quantum states is considered in the
one-loop approximation of quantum cosmology. A universal
effective action
algorithm for the distribution function of chaotic inflationary
cosmologies is derived for both of these states.
The energy scale of inflation is calculated by finding a
sharp probability peak in this distribution function for a
tunnelling model driven by the inflaton field with large negative
constant $\xi$ of nonminimal interaction. The sub-Planckian
parameters of this peak (the mean value of the corresponding Hubble
constant
${\mbox{\boldmath $H$}}\simeq 10^{-5}m_P$, its quantum width
$\Delta{\mbox{\boldmath $H$}}/{\mbox{\boldmath $H$}}\simeq 10^{-5}$
and the number of inflationary
e-foldings ${\mbox{\boldmath $N$}}\simeq 60$) are found
to be in good correspondence with the observational status of
inflation theory,
provided the coupling constants of the theory are constrained by a
condition
which is likely to be enforced by the (quasi) supersymmetric nature
of the
sub-Planckian particle physics model.
\end{abstract}

\baselineskip6.8mm
\section{Introduction}
\hspace{\parindent}
It is widely reckognized that one of the most promising pictures of
the early
universe is a chaotic inflationary scenario \cite{Linde}.
The
inflation
paradigm is the more so attractive that it allows to avoid the
fortune telling
of quantum gravity and cosmology because the inflationary epoch is
supposed to
take place at the energy scale or a characteristic value of the
Hubble constant
$H=\dot a/a \sim 10^{-5}m_P$ much below the Planck one $m_P=G^{1/2}$.
The predictions of  the inflation theory essentially depend on this
energy
scale which must be chosen to provide a sufficient number of
e-foldings $N$ in the expansion law of a scale factor $a(t)$ during
the
inflationary epoch, $N=\int _{t_{\rm in}}^{t_{\rm fin}} dt\,H={\rm
ln}\,(a_{\rm
fin}/a_{\rm in})$,
and also generate the necessary level of density
perturbations capable of the formation of the large scale structure.
In the
chaotic inflationary model the Hubble constant  $H=H(\phi)=\sqrt{8\pi
U(\phi)/3m_P^2}$ is effectively generated by the potential $U(\phi)$
of the
inflaton scalar field $\phi$ which satisfies the slow-roll
approximation
\cite{Linde},
        $\dot\phi\simeq -(1/3H)\partial U/\partial\phi\ll H\phi$.
The number of e-foldings $N=N(\phi_I)$, the effective Hubble constant
$H=H(\phi_I)$ and the generation of the density perturbations, as
well as the
validity of the slow-roll approximation itself,  essentially depend
upon one
parameter -- the initial value of the inflaton field $\phi_I$, and
one of the
most fundamental observational bounds is the following restriction on
this
quantity \cite{Linde}
\begin{equation}
N(\phi_I)\simeq (4\pi/m_P^2)\int_{0}^{\phi_I}d\phi\,H(\phi)
        \Big[\partial H(\phi)/\partial\phi\Big]^{-1} \geq 60 .
\end{equation}
This quantity, however, is a free parameter in the inflation theory,
and, to
the best of our knowledge, there are no convincing principles that
could fix it
 without invoking the ideas of quantum gravity and cosmology.  These
ideas
imply that there exists a quantum state of coupled gravitational and
matter
fields, which in the semiclassical regime generates the family of
inflationary
universes with different values of $H(\phi)$, approximately evolving
at later
times according to classical equations of motion. This quantum state
allows one
to calculate the distribution function $\rho(\phi)$ of these
universes and
interprete its maximum at certain value of $\phi=\phi_I$ (if any) as
generating
the quantum scale of inflation. The implementation of this approach,
undertaken
in the tree-level approximation for the no-boundary \cite{HH,H} and
tunnelling
\cite{Vilenkin} quantum states of  the Universe, was not successful.
The
corresponding distribution functions turned out to be extremely flat
\cite{H-Page,Vilenkin:tun-HH} for large values of $\phi$ (in the
domain of the
inflationary slow-roll ansatz) and unnormalizable at
$\phi\rightarrow \infty$, which totally breaks the
validity of the semiclassical expansion underlying the inflation
theory, since
the contribution of the
over-Planckian energy scales is not suppressed to zero.  Apart from
this
difficulty,  the only local maximum of $\rho(\phi)$ found for the
no-boundary
quantum state was shown to be generating insufficient amount of
inflation
violating the above bound \cite{GrisR}.

In the series of recent papers \cite{BKam:norm,BarvU,tunnel,reduct} we
proposed the
mechanism of suppressing the over-Planckian energy scales by the
contribution
of the quantum (loop) part of the gravitational effective action
to the distribution function of the above type.
This can
justify the use of the semiclassical expansion and serve as a
selection
criterion of physically viable particle models with the normalizable
quantum
state, suggesting the supersymmetric extension of field models in the
theory of the early universe \cite{Kam:super}. In \cite{PhysLett} this
mechanism has been further applied to show that it can also generate
the quantum scale of inflation and, in particular, serve as a quantum
gravitational ground for the inflationary
model of Bardeen, Bond and Salopek \cite{SalopBB} with large negative
constant $\xi$ of nonminimal inflaton-graviton coupling. In this
paper we discuss the result of \cite{PhysLett} with a special emphasis on
the universality of the effective action algorithm for the above
distribution function in both no-boundary and
tunnelling quantum states. We also dwell on the advantages of the
tunnelling proposal for the cosmological wavefunction, that arise both
at the level of applications in the theory of the early Universe and at
the conceptual level of the prospects for the third quantization of
gravity.

\section{Tree-level approximation: nonminimal inflaton field}
\hspace{\parindent}
Two known proposals for the cosmological quantum state, which
semiclassically
generate the families of inflationary universes,  are represented by
the
no-boundary \cite{HH,H} and tunnelling \cite{Vilenkin} wavefunctions.
They
describe two different types of nucleation of the Lorentzian
quasi-DeSitter
spacetime from its Euclidean counterpart  which in the context of
spatially
closed cosmology can be represented by the 4-dimensional Euclidean
hemishere
matched across the equatorial section to the Lorentzian expanding
Universe. The
tree-level no-boundary $\rho_{N\!B}(\phi)$ \cite{HH} and
tunnelling $\rho_{T}(\phi)$ \cite{rhoT} distribution functions of
such
universes are just the squares of their wavefunctions
        \begin{eqnarray}
        \rho_{N\!B}(\phi)\sim {\rm e}^{\textstyle -I(\phi)},\,\,\,
        \rho_{T}(\phi)\sim {\rm e}^{\textstyle +I(\phi)} ,
\label{2.1}
        \end{eqnarray}
where $I(\phi)$ is a doubled Euclidean action of the theory
calculated on such
a hemisphere (or the action on the full quasi-spherical manifold).
When $\phi$
belongs to the domain of the slow-roll approximation and is
practically
constant in the solution of both Lorentzian and Euclidean equations
of motion,
the Euclidean spacetime only slightly differs from the exact
4-dimensional
sphere of the radius  $1/H(\phi)$ -- the
inverse of the
Hubble constant -- and  $I(\phi)$ takes the form
\begin{equation}
        I(\phi)=-3m_P^4/8U(\phi).
\end{equation}
Thus,  $\rho_{NB}(\phi)$ and  $\rho_{T}(\phi)$ describe opposite
outcomes of
the most probable underbarrier penetration: respectively to the
minimum and to
the maximum of the inflaton potential $U(\phi)\geq 0$ (although, in
the former
case the minimum $U(\phi)=0$ generally falls out of the slow-roll domain).

The equations above apply to the case of an inflaton field minimally
coupled to
the metric tensor $G_{\mu\nu}$ with the Lagrangian
\begin{equation}
L(G_{\mu\nu},\phi)=G^{1/2}\left\{\frac{m_P^2}{16\pi}R\,(G_{\mu\nu})-
\frac 12 (\nabla\phi)^2-U(\phi)\right\},
\end{equation}
but can also be used in the theory of
the nonminimal scalar field $\varphi$
        \begin{eqnarray}
        {\mbox{\boldmath $L$}}(g_{\mu\nu},\varphi)
        =g^{1/2}\left\{\frac{m_P^2}{16\pi}R\,(g_{\mu\nu})-
        \frac12 \xi\varphi^2R\,(g_{\mu\nu})
        -\frac 12 (\nabla\varphi)^2
        -\frac12 m^2\varphi^2-\frac{\lambda}{4}\,\varphi^4\right\},
\label{2.3}
        \end{eqnarray}
provided  $L(G_{\mu\nu},\phi)$ above is viewed as the Einstein frame
of ${\mbox{\boldmath $L$}}(\,g_{\mu\nu},\varphi\,)$ with the fields
$(G_{\mu\nu},\phi)=((1+8\pi|\xi|\varphi^2/m_P^2)g_{\mu\nu},\phi(\varphi))$
related to $(\,g_{\mu\nu},\varphi\,)$ by the transformation that
can be found in \cite{SalopBB,Page:conf,BKK}.
For a
negative nonminimal coupling constant $\xi=-|\xi|$  this model easily
generates
the chaotic inflationary scenario \cite{Spok-Unr} with the Einstein
frame
potential
        \begin{eqnarray}
        U(\phi)\,\Big|_{\,\textstyle\phi\!=\!\phi(\varphi)}=
        \frac{m^2\varphi^2/2+\lambda\varphi^4/4}
        {\Big(1+8\pi |\xi\,|\varphi^2/m_P^2\Big)^2} ,
      \label{2.4}
        \end{eqnarray}
including the case of a symmetry breaking at scale $v$ with
$m^2=-\lambda v^2<0$ in the Higgs potential
$\lambda(\varphi^2-v^2)^2/4$ . At large $\varphi$ it approaches
a constant and depending on the parameter
$\delta\equiv -8\pi|\xi|m^2/\lambda m_P^2=8\pi|\xi|v^2/m_P^2$ has two
types of
behaviour at the intermidiate values of the inflaton field. For
$\delta>-1$ it
does not have local maxima and generates the slow-roll decrease of
the
scalar field leading to a standard scenario with a finite
inflationary stage,
while for $\delta<-1$ it has a local maximum at
$\bar\varphi=m/\sqrt{\lambda|1+\delta|}$ and due to a negative slope
of the
potential leads to the eternal inflation for all models with the
scalar field
growing from its initial value  $\varphi_I>\bar\varphi$.

The tree-level distribution functions (\ref{2.1}) for such a
potential do not
suppress the over-Planckian scales and are unnormalizable at large
$\varphi$,
$\int^{\infty}d\varphi \,\rho_{N\!B,\,T}(\varphi) =\infty$, thus
invalidating
a semiclassical expansion. Only in the tunnelling case
with $\delta<-1$ the distribution $\rho_{T}(\phi)$ has a local peak
at $\bar\varphi$, which could have served as a source of the inflation
energy scale at reasonable sub-Planckian value of the Hubble constant.
However, this peak requires the positive mass of the inflaton field
$m^2>\lambda m_P^2/(8\pi|\xi|)$ which is too large even for reasonable
values $\xi=-2\times10^4$, $\lambda=0.05$ \cite{SalopBB} and formally
generates an infinite duration of the inflationary stage (because the
latter starts from the maximum of the inflaton potential).

\section{Beyond the tree-level theory: no-boundary vs tunnelling
wavefunctions}
\hspace{\parindent}
Beyond the tree-level approximation the distribution function for the
inflaton
field $\varphi$ should be regarded as a diagonal element of the
reduced density matrix of this field
${\rm Tr}_f|\Psi\!\!><\!\!\Psi|$. It can be obtained from the
full quantum state $|\Psi\!\!>=\Psi(\phi,f\,|\,t)$ by tracing out
the rest of the degrees of freedom $f$
        \begin{eqnarray}
        \rho\,(\phi\,|\,t)=
        \int df\;\Psi^*(\phi,f\,|\,t)\,\Psi(\phi,f\,|\,t),
\label{3.1}
        \end{eqnarray}
which does not reduce to a simple squaring of the wavefunction (we
begin this section by considering again the minimally coupled inflaton
field). For the inner
product in (\ref{3.1}) to be unambiguously defined, the wavefunction
$\Psi(\phi,f\,|\,t)$ should be taken in the representation of
physical (ADM)
variables with the time $t$ fixed by a chosen ADM reduction procedure
\cite{ADM}. Strictly speaking this reduction is not generally
(globally on phase space of the theory) consistent, and a complete
understanding and the interpretation of the cosmological
wavefunction might be reached only in the framework of the third
quantization of gravity theory. Although this framework still
does not have a status of a well-established physical theory,
there exists a good correspondence principle of this
framework with the quantization in reduced phase space for systems
with a wide class of special (positive-frequency) semiclassical
quantum states. For these states the conserved current of
the Wheeler-DeWitt equations perturbatively coincides with
the inner product of the ADM quantization mentioned above and
thus can be used for the construction of the probability
distribution (for a perturbative equivalence of the ADM and
Dirac-Wheeler-DeWitt
quantization of gravity for such physical states see
\cite{BPon,BKr,BarvU}). As we shall see below, the tunnelling
wavefunction belongs to such a class of states, while the no-boundary
one does not and should be supplied with additional (third quantization)
principles to be interpreted in terms of the probability distribution
of the above type.

In the
approximation of the
Robertson-Walker model,  the ADM physical variables describing a
spatially homogeneous background and inhomogeneous field modes (treated
perturbatively) are respectively the inflaton field $\phi$
and linearized transverse (and traceless) modes $f$ of all possible
spins
\cite{BKam:norm,BarvU,tunnel}, while $t$ can be chosen to be a cosmic
time with
the unit lapse or a conformal time with the lapse $N=a$. In this
approximation the semiclassical solution of the Wheeler-DeWitt equations
can be given by the linear superposition of the two (decaying and
growing) wavefunctions in the underbarrier (Euclidean) regime
$a \leq 1/H$
\begin{equation}
\Psi_{\pm}(a,\phi,f)=\frac 1{[a^2(1-H^2a^2)]^{1/4}}\,e^{\pm I(a,\phi)}
\prod_n\frac 1{\sqrt{u_n}}e^{\pm \frac 12 a^k(\dot{u}_{n}/u_n)
f_n^2},
\end{equation}
\begin{equation}
I(a,\phi)=-\frac{\pi m_P^2}{2H^2}[1-(1-H^2a^2)^{3/2}],\,\,\,
H^2=\frac{8\pi U(\phi)}{3 m_P^2}
\end{equation}
and outgoing and ingoing wavefunctions in the classically allowed
(Lorentzian) regime $a>1/H$ \cite{Wada,VilVach}
\begin{equation}
\Psi^L_{\pm}(a,\phi,f)=\frac 1{[a^2(H^2a^2-1)]^{1/4}}\,e^{\pm iS(a,\phi)}
\prod_n\frac 1{\sqrt{v_n}}e^{\pm \frac i2 a^k (\dot{v}_{n}/v_n)
f_n^2},
\end{equation}
\begin{equation}
S(a,\phi)=-\frac{\pi m_P^2}{2H^2}(H^2a^2-1)^{3/2}.
\end{equation}
Here $I$ and $S$ are the Euclidean and Lorentzian Hamilton-Jacobi functions
of a spatially homogeneous superspace background, the products over $n$
denote the quadratic contribution into these Hamilton-Jacobi functions of the
spatially inhomogeneous modes $f_n$ enumerated by the collective index $n$.
The functions $u_n$ and $v_n$ are their Euclidean and Lorentzian
basis functions
respectively in the semiclassical Euclidean and Lorentzian times defined
according to the classical equations for the minisuperspace background
\begin{equation}
\frac{\partial I}{\partial a}=-\frac{3 m^{2}_{P}}{4 \pi} \frac{a
\dot a}{N}
\end{equation}
(a similar equation holds for Lorentzian time
with $S$ replacing $I$, and for brevity we denote by the dot the
derivatives with respect to both Euclidean $\tau$ and Lorentzian $t$
times). In case of gravitons and minimally interacting massless
scalar particles, the functions $v_n$ satisfy the wave equation
\begin{equation}
\stackrel{..}v_n+k\frac{\dot{a}}a \dot{v_n}
+\frac{n^2-1}{a^{2k-4}} v_n = 0,
\end{equation}
where $n=1,2,...$ for a scalar, $n=3,4,...$ for a graviton and $k$
corresponds to the choice of lapse $N=a^{3-k}$ ($k=3$ for cosmic time
and $k=2$ for a conformal one). Euclidean basis functions $u_n$
satisfy a similar equation with the negative sign of the potential
term.

It is important that the signs of $I$ and $S$ in (3.2) and
(3.4) are correlated with the signs of $f^2$ terms in the
exponentials. Therefore, to provide the normalizability of the
wavefunctions in the space of $f$,
\begin{eqnarray}
Re\left(\pm\frac{\dot {u_n}}{u_n}\right)<0,\,\,\,
Re\left(\pm i\frac{\dot {v_n}}{v_n}\right)<0,    \label{3.4a}
\end{eqnarray}
for different $(\pm)$ branches of the quantum state one should
choose different basis functions among the two independent
solutions of eq.(3.7) and its Euclidean analogue: $v_n^{\pm}$
and $u_n^{\pm}$. In the branch of the growing underbarrier
wavefunction $\Psi_{-}$ (which is just the case of a pure no-boundary
state of Hartle and Hawking) this uniquely leads to the regularity of
$u_{n}^{-}$ on the Euclidean section of the spacetime background, while
for the decaying wavefunction $\Psi_{+}$ (the dominant contribution
to the tunnelling state) $u_{n}^{+}$ is singular at the pole of the
Euclidean hemisphere \cite{VilVach}.

The wavefunctions (3.2) and (3.4) are the building blocks
of the semiclassical tunnelling and no-boundary wavefunctions.
The no-boundary wavefunction prescribed by the Euclidean path integral
turns out to be the following in the Euclidean and Lorentzian
regimes \cite{VilVach}
\begin{equation}
\Psi_{NB}=\Psi_{-},\,\, a<1/H,  \label{3.5a}
\end{equation}
\begin{equation}
\Psi_{NB}=e^{\pi m_P^2/2 H^2}\left[e^{i\pi/4}\Psi_{+}^L+
e^{-i\pi/4}\Psi_{-}^L\right],\,\, a>1/H,  \label{3.6a}
\end{equation}
while the tunnelling wavefunction $\Psi_{T}$ in the prescription of
the outgoing wave of ref.\cite{Vilenkin:tun-HH} looks as follows
(the overall normalization of $\Psi_{T}$ is fixed by the requirement
that $\Psi_{T}$ is $\phi$-independent at $a = 0$):
\begin{equation}
\Psi_{T}=e^{-\pi m_P^2/2 H^2}\Psi_{+}^{L},\,\, a>1/H,\label{3.7a}
\end{equation}
\begin{equation}
\Psi_{T}=\Psi_{+}-\frac{i}2
e^{-\pi m_P^2/ H^2}\Psi_{-},\,\, a<1/H.     \label{3.8a}
\end{equation}

As shown in \cite{Vilenkin:tun-HH,VilVach}, the requirement of the
normalizability
of the wavefunction in $f$ (\ref{3.4a}) (in both branches of
the underbarrier wavefunction (\ref{3.8a})) and the matching conditions
at the nucleation point $a=1/H$ uniquely singles out $v_n^{+}(t)$ in
(\ref{3.7a}) to be the negative frequency basis function of the
DeSitter invariant vacuum in the Lorentzian expanding Universe.
This differs from the proposal in ref.\cite{Rubak} where the vacuum
state was specified at $a=0$ instead of $a > 1/H$ (see also
\cite{Wada}).
In the conformal time $t$ $(k=2)$, in which the DeSitter Lorentzian
background has the form
\begin{eqnarray}
a=(H \cos t)^{-1},\,\, 0\leq t<\pi/2,
\end{eqnarray}
this basis function is given by
\begin{eqnarray}
v_n^{+}(t)=\frac{(z-1)^{(n-1)/2}}{(z+1)^{(n+1)/2}}\left(1+
\frac{z}{n}\right),\ \ z=-i \tan t.
\end{eqnarray}
The
Euclidean basis functions $u_n^{\pm}(\tau)$ in (\ref{3.8a}) are the
analytic continuation
\begin{eqnarray}
u_n^{\pm}(\tau)=v_n^{+}(\mp i\tau),\  0\leq\tau<\infty,
\end{eqnarray}
to the imaginary values of the Lorentzian time. The analytic
continuation $t=i\tau$ matches at $\tau=t=0$ the Lorentzian
DeSitter Universe with
the Euclidean DeSitter hemisphere having the scale factor
\begin{eqnarray}
a=(H \cosh \tau)^{-1}, 0\leq \tau<\infty.
\end{eqnarray}
Note that $u_n^{+}(\tau)$ is singular at the pole of this hemisphere
$\tau=+\infty$, while $u_n^{-}(\tau)$ is regular there. On the contrary,
these two functions are correspondingly regular and singular at
the point $\tau=-\infty$ which can be regarded as a pole of the hemisphere
$-\infty<\tau\leq 0$ complimentary to the above one on the full
4-dimensional sphere -- the DeSitter gravitational instanton
obtained by glueing the Euclidean hemisphere with its double
\cite{BKam:norm,tunnel}.

Now we can calculate the probability distribution
$\rho\,(\phi\,|\,t)$ for the tunnelling and
no-boundary quantum states. This distribution makes sense only in
the Lorentzian domain $a>1/H$ and in the tunnelling case
consists in squaring the single outgoing component (3.12),
dropping the first preexponential factor in eq.(3.4) (this
corresponds to calculating the needed conserved current of the
Wheeler-DeWitt equation or the reduction to the physical state
of the ADM quantization \cite{BKr}) and taking the gaussian integral
over $f$. The result looks as follows
\begin{equation}
\rho\,(\phi\,|\,t)=
\prod_n \, [\Delta_n]^{-1/2}\,
e^{-\pi m_P^2/H^2},
\end{equation}
\begin{equation}
\Delta_n=ia^k(v_n^{+}\dot{v}_n^{+*}
-v_n^{+*}\dot{v}_n^{+}).
\end{equation}
The preexponential factor here is given by the product of the Wronskians
of the Lorentzian basis functions $v_n^{+}$ and $v_n^{+*}$, which are
$t$-independent and can be calculated at the nucleation point $t=0$.
As it was shown in \cite{BKam:norm,BarvU,tunnel,reduct} this product coincides
with the product
of Wronskians of the pairs of Euclidean basis functions $u_n^{\pm}$
regular on the opposite (complimentary) hemispheres of the DeSitter
gravitational instanton and, actually, comprises the exponentiated
one-loop Euclidean effective action of the full set of fields $f$,
$\SGamma_{\rm 1-loop}(\phi)=(1/2)\sum_n {\rm ln} \Delta_n$
so that the probability distribution takes the form
\begin{eqnarray}
        \rho_{T}(\phi)\cong
         \frac{1}{H^{2}(\phi)}\;{\rm e}^
        {\,I(\phi)-
        \SGamma_{\rm 1-loop}(\phi)}.
        \end{eqnarray}
Here $I(\phi)=-\pi m_P^2/H^2(\phi)$
is the classical action on the
DeSitter instanton, the effective action is chosen
to include also the contribution
of the quantum inflaton mode (which is compensated by additional
multiplyer $1/H^2 (\phi)$) and can be
represented in the form of the functional trace of the logarithm of
the inverse propagator of the full system of fields $g(x)$
inhabiting the Universe
\begin{eqnarray}
\SGamma_{\rm 1-loop}(\phi)=
        \left.\frac12\,{\rm Tr\,ln}\,
        \frac{\delta^2 I[\,g\,]}{\delta g(x)\,\delta g(y)}
        \,\right |_{\,\rm \bf D\!S}.
\end{eqnarray}
It is calculated on the quasi-DeSitter gravitational instanton {\bf
DS} -- the
4-dimensional quasi-sphere of radius $1/H(\phi)$ --
and,
therefore, parametrized by $\phi$.

In the case of the no-boundary state, the Lorentzian wavefunction
consists of the outgoing and ingoing components which describe the
expanding and contracting Universes. Therefore, the reduced phase space
quantization does not work, as well as the current of the Wheeler-DeWitt
equation turns out to be zero in view of the reality of the wavefunction.
Therefore, the only hope to interprete this situation is to invoke certain
ideas of the third quantization and regard the wavefunction as describing
the two coexisting Universes propagating in opposite directions in the
minisuperspace of the scale factor. Then the calculation of the
probability distribution of every of such Universes heuristically
implies projecting the full wavefunction on one of the
components and then repeating the same calculations that obviously
lead to the same algorithm in terms of the effective action of the theory
on the DeSitter instanton. The fact that the regular and singular
Euclidean modes get interchanged as compared to the tunnelling case
does not change the result because they both symmetrically enter the final
algorithm.

Thus the probability distribution for no-boundary
and tunnelling quantum states can be both represented in one
equation and applied to the case of the nonminimal inflaton field
\cite{BKam:norm,BarvU,tunnel,PhysLett}
        \begin{eqnarray}
        {\mbox{\boldmath $\rho$}}_{N\!B,\,T}(\varphi)\cong
         \frac1{{\mbox{\boldmath $H$}}^2 (\varphi)}\;{\rm e}^
        {\textstyle\, \mp {\mbox{\boldmath $I$}}(\varphi)-
        {\mbox{\boldmath $\Gamma$}}_{\rm 1-loop}(\varphi)},
\label{3.2}
        \end{eqnarray}
where ${\mbox{\boldmath $I$}}(\varphi)=I(\phi(\varphi))$ is the
Euclidean
action rewritten in the frame of the original Lagrangian
(\ref{2.3}), ${\mbox{\boldmath $H$}}(\varphi)$ is a Hubble constant
in the same
frame related by the equation
\begin{equation}
{\mbox{\boldmath
$H$}}(\varphi)=H(\phi)\sqrt{1+8\pi|\xi|\,\varphi^2/m_P^2}
\end{equation}
to the Hubble constant $H(\phi)$ in the Einstein frame and
${\mbox{\boldmath $\Gamma$}}_{\rm 1-loop}(\varphi)$ is the effective action
in the original field frame (remember that we denote the quantities
in this frame by boldface letters to distinguish them
from those of the Einstein one). Note that in contrast to the classical
action the effective actions calculated in different field frames are
numerically different and do not differ only by the redefinition
of their field argument (even on shell). This would be the case for
the local
reparametrizations of fields only up to the renormalization
procedure which induces the dimensional cutoff defined relative to
a given parametrization of the spacetime metric. In what follows we
consider as physical (that is defining the renormalization scale) the
metric $g_{\mu\nu}$ in the original parametrization, which means that
${\mbox{\boldmath $\Gamma$}}_{\rm loop}(\varphi)$ should be defined
and renormalized in the original field frame. This is of crucial
importance for obtaining the correct scaling behaviour of the
effective action and the probability distribution.

Indeed, in the high-energy limit of the large inflaton field,
including the slow-roll domain
and corresponding in the model (\ref{2.3}) to the
Hubble constant ${\mbox{\boldmath $H$}}(\varphi)\simeq
\sqrt{\lambda/12|\xi|}\varphi\rightarrow\infty$,
the effective action is calculated and renormalized
on the DeSitter instanton
of vanishing {\it physical} size ${\mbox{\boldmath $H$}}^{-1}$ and,
therefore, is determined asymptotically by the total anomalous
scaling $Z$ of the theory on such a manifold
        \begin{eqnarray}
        {\mbox {\boldmath$ \Gamma$}}_{\rm 1-loop}\,\Big|
        _{\,\textstyle
{\mbox{\boldmath $H$}}
\!\rightarrow\!\infty}
        \simeq Z\,{\rm ln}\,\frac{{\mbox{\boldmath $H$}}}{\mu}.
\label{3.4}
        \end{eqnarray}
Here $\mu$ is a renormalization mass parameter or a dimensional
cutoff
generated by the fundamental and finite string theory, if the model
(\ref{2.3})
is regarded as its sub-Planckian effective limit. This is in sharp
contrast to the renormalization in the unphysical metric $G_{\mu\nu}$
that would have led to the above equation with
${\mbox{\boldmath $H$}}(\varphi)$ replaced by
$H(\phi)\rightarrow\sqrt{\lambda/96\pi \xi^2}m_P$, featuring absolutely
different asymptotic behaviour in $\varphi$.
In the one-loop
approximation
the parameter $Z$ is determined by the total second DeWitt
coefficient
\cite{DW:Dynamical} of all quantum fields  $g=(\varphi,f)$,
integrated over the DeSitter instanton,
\begin{equation}
Z=\frac{1}{16\pi^2}\,\int_{\rm \bf D\!S} d^4x\,g^{1/2}a_2(x,x)
\end{equation}
and, thus, crucially depends on the particle content of a model
including as
a graviton-inflaton sector the Lagrangian (\ref{2.3}). This
quantity, in particular, determines the one-loop divergences
of the field model and the set of corresponding $\beta$-functions.

The use of eqs.(\ref{3.2}) and (\ref{3.4}) shows that the quantum
probability
distribution acquires in contrast to its tree-level approximation
(\ref{2.1})
extra $Z$-dependent factor
        \begin{eqnarray}
        {\mbox{\boldmath $\rho$}}_{N\!B,\,T}(\varphi)\cong
        e^{\pm 3m_P^4/8U(\phi\,(\varphi))}
        \varphi^{-Z-2},
                \label{3.6}
        \end{eqnarray}
which can make the both no-boundary and tunnelling wavefunctions
normalizable
at over-Planckian scales provided the parameter $Z$ satisfies the
inequality
\begin{equation}
Z>-1
\end{equation}
serving as a selection criterion for consistent particle
physics models
with a justifiable semiclassical loop expansion
\cite{BKam:norm,Kam:super}.
Although this equation is strictly valid only in the limit
$\varphi\rightarrow\infty$, it can be used for a qualitatively good
description
at intermidiate energy scales. In this domain the distribution
(\ref{3.6}) can
generate the inflation probability peak at $\varphi=\varphi_I$ with
the
dispersion
$\sigma$, $\sigma^{-2}=-d^2 {\rm ln}\,{\mbox{\boldmath
$\rho$}}(\varphi_I)/
d\varphi_I^2$,
        \begin{eqnarray}
        \varphi_I^2=\frac{2|I_1|}{Z+2},\,\,\,\,\sigma^2=\frac{|I_1|}{(
Z+2)^2},\,\,\,\,
        I_1=-24\pi\frac{|\xi|}{\lambda}\,(1+\delta)\,m_P^2,
       \label{3.7}
        \end{eqnarray}
where $I_1$ is a second coefficient of expansion of the Euclidean
action in
inverse powers of $\varphi$, $ {\mbox{\boldmath $I$}}(\varphi)=
-3m_P^4/8U(\phi\,(\varphi))=I_0+I_1/\varphi^2+O(1/\varphi^4)$. For
the
no-boundary and tunnelling states this peak exists in complimentary
ranges of the parameter $\delta$. For the no-boundary state it can be
realized
only for $\delta<-1\,(I_1>0)$ and, thus, corresponds to the endless
inflation with the field $\varphi$ on the negative slope of the
inflaton potential (\ref{2.4}) growing from its starting value
$\varphi_I>\bar\varphi$.
For a tunnelling proposal this peak takes place for $\delta\!>\!-1$
and
generates
the finite duration of the inflationary stage with the number of
e-foldings in
the original frame
        ${\mbox{\boldmath $N$}}(\varphi_I)=
        (\varphi_I/m_P)^2
\pi(|\xi|+1/6)/(1+\delta)=8\pi^2|\xi|(1+6|\xi|)/\lambda(Z+2)$.
In what follows we consider the latter case because it describes the
conventional scenario with the matter-dominated stage following the
inflation.

\section{Nonminimal inflation and particle physics of the early
Universe}
\hspace{\parindent}
The status of the inflation theory has recently been
strongly confirmed by the observations of the
cosmic microwave background radiation anisotropy in
the COBE \cite{COBE} and Relikt \cite{Relikt} satellite experiments.
In the chaotic inflationary model with a nonminimal inflaton
field (\ref{2.3}) the spectrum of perturbations compatible
with these measurements can be acquired in the range of
coupling constants $\lambda/\xi^2\sim 10^{-10}$ \cite{SalopBB,Salopek}
(the experimental
bound on the
gauge-invariant \cite{BST} density perturbation $P_{\zeta}(k)=N^2_k
(\lambda/\xi^2)/8\pi^2$ in the $k$-th mode "crossing" the horizon at
the moment
of the e-foldings number $N_k$).  The main advantage of this model is that
it allows one to avoid an
unnaturally
small value of $\lambda$ in the minimal inflaton model \cite{Linde}
and replace
it with the GUT compatible value $\lambda\simeq 0.05$, provided
$\xi\!\simeq\!
-2\!\times\! 10^4$ is chosen to be related to the ratio of the Planck
scale to a
typical GUT scale, $|\xi|\sim m_P/v$. For these coupling constants
the bound
${\mbox{\boldmath $N$}}(\varphi_I)\geq 60$ on the duration of the
inflation, generated by the probability peak (\ref{3.7}), results in
an
enormous value of the anomalous scaling $Z\sim 10^{11}$. A remarkable
feature
of the proposed scheme is that this huge value can be naturally
induced by
large $\xi$ already in the one-loop approximation. Indeed, the
expression for
$Z_{1-\rm loop}$, well known for a generic theory
\cite{DW:Dynamical}, has a
contribution quartic in effective masses of  physical particles
easily
calculable on a spherical DeSitter background \cite{Al-FrTs}
\begin{equation}
Z_{1-\rm loop}=(12{\mbox{\boldmath
$H$}}^4)^{-1}(\sum_{\chi}m_{\chi}^4
+4\sum_{A}m_{A}^4-4\sum_{\psi}m_{\psi}^4)+...,
\end{equation}
where the summation goes over all Higgs scalars $\chi$, vector gauge
bosons $A$
and Dirac spinors $\psi$. Their effective masses for large $\varphi$
are dominated by the contributions
\begin{equation}
m_{\chi}^2=\lambda_{\chi}\varphi^2/2,\,\,m_{A}^2=g_{A}^2\varphi^2,\,\,
m_{\psi}^2=f_{\psi}^2\varphi^2
\end{equation}
induced via the Higgs mechanism from
their
interaction Lagrangian with the inflaton field
        \begin{eqnarray}
        {\mbox{\boldmath $L$}}_{\rm
int}=\sum_{\chi}\frac{\lambda_{\chi}}4
        \chi^2\varphi^2
        +\sum_{A}\frac12 g_{A}^2A_{\mu}^2\varphi^2+
        \sum_{\psi}f_{\psi}\varphi\bar\psi\psi + {\rm derivative\,\,
coupling}.
\label{4.1}
        \end{eqnarray}
Thus, in view of the relation $\varphi^2/{\mbox{\boldmath
$H$}}^2=12|\xi|/\lambda$, we get the leading contribution of large
$|\xi|$ to
the total anomalous scaling of the theory
        \begin{equation}
        Z_{\rm 1-loop}=6\,\frac{\xi^2}{\lambda}{\mbox{\boldmath
$A$}}+O(\xi),
        \end{equation}
        \begin{equation}
        {\mbox{\boldmath
$A$}}=\frac1{2\lambda}\Big(\sum_{\chi}\lambda_{\chi}^2
        +16\sum_{A}g_{A}^4-16\sum_{\psi}f_{\psi}^4\Big) ,
\label{4.2}
        \end{equation}
which contains the same large dimensionless ratio
$\xi^2/\lambda\simeq 10^{10}$
and the universal quantity ${\mbox{\boldmath $A$}}$ determined by a
particle
physics model (gravitons and inflaton field do not contribute to
${\mbox{\boldmath $A$}}$, as well as gravitino in case when the
latter is
decoupled from the inflaton).

For such $Z_{\rm 1-loop}$ the parameters of the inflationary peak
express as
        \begin{equation}
        \varphi_I=m_P\sqrt{\frac{8\pi(1+\delta)}{|\xi|\,
        {\mbox{\boldmath $A$}}}},\,\,\,\,
        \sigma=\frac{\varphi_I}{\sqrt{12{\mbox{\boldmath $A$}}}}
        \frac{\sqrt{\lambda}}{|\xi|},
        \end{equation}
        \begin{equation}
        {\mbox{\boldmath $H$}}(\varphi_I)=
        m_{P}\frac{\sqrt{\lambda}}{|\xi|} \sqrt{\frac{2\pi(1+\delta)}
{3A^{2}}},
\,\,\,\,
        {\mbox{\boldmath $N$}}(\varphi_I)=
        \frac{8\pi^2}{\mbox{\boldmath $A$}}
\label{4.3}
        \end{equation}
and satisfy the bound  ${\mbox{\boldmath $N$}}(\varphi_I)\geq 60$
with a single
restriction on ${\mbox{\boldmath $A$}}$, ${\mbox{\boldmath
$A$}}\leq1.3$.
This restriction
justifies a slow-roll approximation, because
the
corresponding smallness parameter (in the original frame of the
Lagrangian
(\ref{2.3})) is
$\dot\varphi/{\mbox{\boldmath $H$}}\varphi\simeq-
{\mbox{\boldmath $A$}}/96\pi^2
\sim-10^{-3}$. For a value of $\delta \ll1$ ($\delta\sim 8\pi/|\xi|$
for
$|\xi|\sim m_P/v$) and ${\mbox{\boldmath $A$}}\simeq 1$,  the
obtained
numerical parameters describe extremely sharp inflationary peak at
$\varphi_I$ with small width and sub-Planckian Hubble constant
\begin{equation}
\varphi_I\simeq 0.03 m_P,\,\,
\sigma\simeq 10^{-7}m_P,\,\,{\mbox{\boldmath
$H$}}(\varphi_I)\simeq
10^{-5}m_P,
\end{equation}
which is the most realistic range of the
inflationary
scenario.  The smallness of the width does not, however, lead to its
quick
quantum spreading: the commutator relations for operators
${\hat\varphi}$ and
$\dot{\hat\varphi}$, $[{\hat\varphi},\dot{\hat\varphi}]\simeq
i/(12\pi^2|\xi|a^3)$ \cite{SalopBB}, give rise at the beginning of
the
inflation, $a\simeq {\mbox{\boldmath $H$}}^{-1}$, to a negligible
dispersion of
$\dot\varphi$, $\Delta\dot\varphi\simeq {\mbox{\boldmath
$H$}}^3/12\pi^2|\xi|\sigma
\simeq(8/{\mbox{\boldmath
$A$}})(\sqrt{\lambda}/|\xi|)|\dot\varphi|\ll|\dot\varphi|$. It is
remarkable
that the relative width
\begin{eqnarray}
\sigma/\varphi_I\sim\Delta{\mbox{\boldmath $H$}}
/{\mbox{\boldmath $H$}}\sim 10^{-5}
\end{eqnarray}
corresponds to the observable
level of
density perturbations, although it is not clear whether this quantum
dispersion
$\sigma$ is directly measurable now, because of the stochastic noise
of the
same order of magnitude generated during the inflation and
superimposed upon
$\sigma$.

All these conclusions are rather universal and (apart from the choice
of
$|\xi|$ and $\lambda$) universally depend on one parameter
${\mbox{\boldmath
$A$}}$ (\ref{4.2}) of the particle physics model.  This quantity
should satisfy
the bound
        \begin{eqnarray}
        0<{\mbox{\boldmath $A$}}\leq 1.3              \label{4.4}
        \end{eqnarray}
in order to render $Z$ positive, thus suppressing over-Planckian
energy
scales, and  provide sufficient amount of inflation
(${\mbox{\boldmath $A$}}$
should not, certainly, be exceedingly close to zero, not to suppress
the
dominant
contribution of large $|\xi|$ in (\ref{4.2})). This bound again
suggests the
quasi-supersymmetric nature of the particle model, although for
reasons
different from the conclusions of
\cite{Kam:super}.  It is only supersymmetry that can constrain the
values of
the Higgs $\lambda_{\chi}$, vector gauge $g_A$ and Yukawa $f_{\psi}$
couplings
so as to provide a subtle balance between the contributions of bosons
and
fermions in (\ref{4.2}) and fit the quantity
${\mbox{\boldmath $A$}}$ into a narrow range (\ref{4.4}).  In
contrast to
ref.\cite{Kam:super}, this conclusion is robust against the
subtleties of the
definition of $Z$ (related to the treatment of zero modes on DeSitter
background \cite{Al-FrTs}) because it probes only the large limit of
$Z\gg 1$.

\section{Conclusions}
\hspace{\parindent}
Thus, the same mechanism that suppresses the over-Planckian
energy scales also generates a narrow probability
peak in the distribution of tunnelling inflationary universes and
strongly
suggests the (quasi)supersymmetric nature of their particle content.
It
seems to be consistent with microwave background
observations
within the model with a strongly coupled nonminimal inflaton field.
A remarkable feature of this result is that it is mainly based on one
small parameter -- the dimensionless ratio of two major energy scales,
the GUT and Planck ones, given by the combination of the coupling
constants $\sqrt{\lambda}/|\xi|\simeq 10^{-5}$.
This result
is
independent of the renormalization ambiguity, which gives a hope that
it is
also robust against  inclusion of multi-loop corrections. It is usual
to be
prejudiced against a large value of the nonminimal coupling $|\xi|$
which
generates
large quantum effects leaving them uncontrollable in multi-loop
orders.  This
is not, however, quite correct, because the effective gravitational
constant in
such a model is inverse proportional to $m_P^2+8\pi|\xi|\varphi^2$
and, thus,
large $|\xi|$ might improve the loop expansion \cite{BKK}. Obviously,
the large
value of $|\xi|$ at sub-Planckian (GUT) scale requires explanation
which might
be based on the renormalization group approach (and its extension to
non-renormalizable theories \cite{BKK}). As shown in \cite{BKK},
quantum
gravity with nonminimal scalar field has an asymptotically free
conformally
invariant ($\xi=1/6$) phase at over-Planckian regime, which is
unstable at
lower energies. It is plausible to conjecture that this instability
can lead
(via composite states of the scalar field) to the inversion of the
sign of
running $\xi$ and its growth at the GUT scale, thus making possible
the
proposed inflation applications \cite{BK}.

As far as it concerns the GUT and lower energy scales,  the ground
for
supersymmetry of the above type looks
very promising in the context of a special property of supersymmetric
models to
have a single unification point for weak, electromagnetic and strong
interactions
(the fact that has been discovered in 1987 and now becoming widely
recognized
after the recent experiments at LEP \cite{supersym}).

 From the viewpoint of the theory of the early universe,
the obtained results give a strong preference to the
tunnelling quantum state. The debate on the advantages of
the tunnelling versus no-boundary wavefunction has a long history
\cite{Linde,Vilenkin,rhoT,Rubak,Vilenkin:tun-HH}. At present,
in the cosmological context the tunnelling
proposal seems to be more useful and conceptually clearer than
the no-boundary one, because for its interpretation one should
not incorporate vague ideas of the third quantization of gravity
which inevitably arise in the no-boundary case: splitting the
Lorentzian wavefunction in positive and negative frequency
parts and separately calculating their probability distributions.
On the other hand, the formulation of the tunnelling proposal is
not so aesthetically closed, for it involves imposing outgoing
condition after the potential barrier, the unit normalization
condition -- before the barrier at $a=0$, the requirement of
the normalizability in variables $f$, etc. And all this in
contradistinction to the closed path-integral formulation of
the no-boundary proposal, automatically providing many of the above
properties. On the other hand, outside of the cosmological
framework, in particular, within the scope of the wormhole and
black hole physics, the tunnelling proposal seems to be helpless.
Moreover, at the overlap of the cosmological framework with the
theory of the virtual black holes it leads to contradictions
signifying that the quantum birth of bigger black holes is more
probable than the small (Planckian) ones \cite{Bousso}.
All these arguments can hardly be
conclusive, because it might as well happen that the difference
between the no-boundary and
tunnelling wavefunctions should be ascribed to the open problem of the
correct quantization of the conformal mode. Note that the
normalizability
criterion for the distribution function  and its algorithm
(\ref{3.2}) do not
extend to the low-energy limit $\varphi\rightarrow 0$, where the
naively
computed no-boundary distribution function blows up to infinity, the
slow-roll
approximation becomes invalid, etc. This is a domain related to a
highly
speculative (but, probably, inevitable) third quantization of gravity
\cite{Coleman}, which goes beyond the scope of this paper.
Fortunately, this
domain is separated from the obtained inflationary peak by a vast
desert with
practically zero density of the quantum distribution, which
apparently
justifies our conclusions disregarding the ultra-infrared physics of
the
Coleman theory of baby universes and cosmological constant
\cite{Coleman}.

\section*{Acknowledgements}
\hspace{\parindent}
The authors benefitted from helpful discussions with Don N.Page,
V.A.Rubakov, D.Salopek, A.A.Starobinsky and A.V.Toporensky. Partly
this work has been made possible
due to the support by the Russian Research Project ``Cosmomicrophysics".
One of the authors (A.O.B.) is grateful for the support of this work
provided by the Russian Foundation for Fundamental Research under
Grant 93-02-15594, International (Soros) Science Foundation and
Government of the Russian Federation Grant MQY300 and the European
Community Grant INTAS-93-493. A.Yu.K. is greateful for the support
of this work provided by the Russian Foundation for Fundamental Research
under
Grant 94-02-03850, International (Soros) Science Foundation and
Government of the Russian Federation Grant MAE300.


\end{document}